\documentstyle[preprint,aps,eqsecnum,floats,epsf,fleqn]{revtex}
\input{acp.com}
\begin{document}
\newfont{\eufxii}{eufm10 scaled 1200}
\newfont{\eusix}{eusm10 scaled 900}
\newfont{\eusxii}{eusm10 scaled 1200}
\newfont{\eurxii}{eurm10 scaled 1200}
\draft
\setcounter{page}{0}
\title{LOCAL ENTROPY CHARACTERIZATION OF CORRELATED RANDOM MICROSTRUCTURES}
\author{
C. Andrau$\mbox{\rm d}^{1}$,
A. Beghdad$\mbox{\rm i}^{2}$,
E. Haslun$\mbox{\rm d}^{3}$,
R. Hilfe$\mbox{\rm r}^{3,4}$\footnote{present address: 
ICA-1, Universit{\"a}t Stuttgart, Pfaffenwaldring 27, 70569 Stuttgart},
J. Lafai$\mbox{\rm t}^{1}$, and 
B. Virgi$\mbox{\rm n}^3$}
\address{
$\mbox{ }^1$Laboratoire d'Optiques des Solides,
Universite Pierre et Marie Curie, CNRS, 4 Place Jussieu, 
75252 Paris Cedex 05, France\\
$\mbox{ }^2$LPMTM-CNRS, Institut Galilee,
Universite Paris Nord, 93430 Villetaneuse, France\\
$\mbox{ }^3$Institute of Physics,
University of Oslo, P.O.Box 1048, 0316 Oslo, Norway\\
$\mbox{ }^4$Institut f{\"u}r Physik,
Universit{\"a}t Mainz,
55099 Mainz,
Germany}
\maketitle
\thispagestyle{empty}
\begin{abstract}
A rigorous connection is established between  the local porosity
entropy introduced by Boger et al. (Physica A {\bf 187}, 55 (1992)) 
and the configurational entropy of Andraud et al. (Physica A {\bf 207},
208 (1994)). These entropies were introduced as morphological
descriptors derived from local volume fluctuations in arbitrary
correlated microstructures occuring in porous media, composites
or other heterogeneous systems.
It is found that the entropy lengths at which the entropies
assume an extremum become identical for high enough resolution
of the underlying configurations.
Several examples of porous and heterogeneous media are given
which demonstrate the usefulness and importance of this morphological local
entropy concept.
\end{abstract}

\section{Introduction}
An accurate  geometric characterization of the correlated random 
microstructures found in disordered heterogeneous materials 
is generally difficult. 
One possible method consists
in using globally well-defined morphological descriptors
(e.g. volume fraction or specific internal surface area) as local
descriptors.
While local volume fraction fluctuations have been discussed 
previously in the context of porous media and photographic granularity
\cite{BB86b,LT90a,one63} the idea of local morphological descriptors 
has only recently been applied systematically in the form of local 
porosity theory 
\cite{hil91d,hil92a,hil92b,hil92f,hil93b,hil93c,hil94b,hil94g,hil95d}.
Local porosity theory generalizes the successful and widely used
effective medium approximation to correlated random microstructures.

Microstructural information of a different kind can be obtained
by applying the Gibbs-Shannon entropy concept from information
theory to distributions of locally fluctuating morphological 
descriptors. 
Such ``local geometry entropies'' were introduced independently
in \cite{hil92f} and \cite{ABL94}.
It is the primary purpose of the present paper to clarify the
similarities and differences between the two different local
geometry concepts. 
Furthermore we resolve the normalization problem of \cite{ABL94},
and provide examples of the usefulness and microstructural
sensitivity of the local geometry concept.
We begin with a discussion of the ``local porosity entropy'' 
introduced in \cite{hil92f} in the next section. 
Subsequently we present the definition of the ``configurational
entropy'' as introduced in \cite{ABL94}.
In section \ref{comparison} we establish the relationship between
the two concepts, and in section \ref{length} we show that, for
sufficiently high resolution,  they give rise to the same 
``entropy length''.
Finally we apply our entropic analysis to several examples
including percolation models, Poisson grain models, thin films
and porous sandstones.

\section{Local porosity entropy}
\label{lpt}
Distributions of local morphological descriptors of random
microstructures were introduced in \cite{hil91d,hil92b}.
The local geometry approach was recently reviewed and 
extended in \cite{hil95d}.
In the present study we consider for simplicity only two
component media in two dimensions.
Such media are of practical importance as granular metal films
for composite coatings.
They arise also as surface morphologies
when sectioning a three-dimensional medium.

Given a two-component random microstructure in the plane $\RR^2$
let the subset $\PP\subset\RR^2$ denote the set occupied by one of 
the two components (phases),  and $\MM=\RR^2\setminus\PP$ its 
complement occupied by the second component. 
For microstructures obtained by sectioning a porous medium 
the set $\PP$ represents the fluid-filled pores while the set $\MM$ 
is the solid mineral matrix.
The local volume fraction occupied by the set $\PP$ within an
observation region $\KK$ is then defined as \cite{hil91d,hil92a,hil95d}
\begin{equation}
\phi(\KK)=\frac{V(\PP\cap\KK)}{V(\KK)}
\label{lp}
\end{equation}
where $V(\GG)$ denotes the volume (=area) of a set $\GG\subset\RR^2$.
The local porosity distribution is defined as the probability
density \cite{hil91d,hil92a,hil95d}
\begin{equation}
\mu(\phi;\KK) = \langle\delta(\phi-\phi(\KK))\rangle
\label{lpd}
\end{equation}
where $\delta(x)$ is the Dirac $\delta$-distribution, and
$\langle...\rangle$ denotes an average over the underlying
probability distribution governing the configurations of the
random microstructure.
Note that the local porosity distribution depends 
on the size and shape of the observation region $\KK$.
Recently \cite{hil95d} it was found that this dependence 
disappears in a suitable macroscopic scaling limit.
In the following it will be assumed that for each $\KK$
the function $\mu(\phi;\KK)$ is a given continuous 
probability density with support on the unit interval.
This assumption can be justified by arguing that possible
discrete components are always smeared because of finite
image resolution.

Local porosity entropies $I(\KK)$ are obtained by calculating
the Gibbs-Shannon entropy of the family of local porosity 
distributions $\mu(\phi;\KK)$ as in \cite{hil92f}
\begin{equation}
I(\KK) = \int_0^1 \mu(\phi;\KK)\log\mu(\phi;\KK)\;d\phi
\label{lpe}
\end{equation}
which assumes, as usual, a uniform distribution as the a priori weight.
Examples of local porosity entropies were given in \cite{hil92f}
for synthetic images, and in \cite{hil94g} for experimental
systems.

\section{``Configuration'' Entropy}
\label{ce}
In \cite{BALPP93,ABL94} the two-dimensional image is discretized into 
black and white picture elements (pixels) forming a quadratic
lattice with lattice constant $a$.
The black and white image is then analyzed using a quadratic
observation region $\KK$ (sliding cell) of sidelength $L$
which contains $M=(L/a)^2$ pixels.
The observation region is moved to $N$ different positions,
and the number $N_k(M)$ is defined as the number of cells of size $M$
containing $k$ active (=black) pixels.
Using the relative frequencies
\begin{equation}
p_k(M) = \frac{N_k(M)}{N}
\end{equation}
as estimators for probabilities the ``configuration''
entropy has been defined as \cite{ABL94}
\begin{equation}
H^*(M)= 
\frac{H(M)}{\log(M+1)} =
-\frac{1}{\log(M+1)}\sum_{k=0}^M p_k(M)\log p_k(M) .
\label{H*}
\end{equation}
where $H(M)$ represents the usual Gibbs-Shannon entropy of the
discrete probabilities $p_k(M)$, and $M=(L/a)^2$.

The expression (\ref{H*}) does not use the Gibbs-Shannon expression
$H(M)$ but divides it with $\log(M+1)$ similar to multiplicative
renormalization in the theory of critical phenomena.
The normalization was introduced because 
the underlying probability space changes when the number
$M$ of pixels inside the measurement cell is changed.
The choice $1/\log(M+1)$ for the normalization 
factor however renders $H^*(M)$ nonadditive. 
The next section will show that this can be avoided 
by using the definition (\ref{lpe}), and that a precise
relation exists between $I(\KK)$ and $H^*(M)$.

\section{Relation between the two entropies}
\label{comparison}

The relationship between $I(\KK)$ and $H^*(M)$ is established
by applying the definitions of section \ref{lpt} to the same
discretized quadratic observation region $\KK$ of sidelength 
$L$ which was used in section \ref{ce}.
For such a choice of $\KK$ the unit interval $[0,1]$ of porosities
is conveniently subdivided into subintervals bounded by the porosities
\begin{equation}
\phi_k=\frac{k}{M+1}
\end{equation}
where $0\leq k\leq M+1$ and $M=(L/a)^2$.
If $\mu(\phi;\KK)$ is given as a continuous function 
the integral in equation (\ref{lpe}) can be approximated as
$I(\KK)=\lim_{M\rai}I(M)$ where 
\begin{eqnarray}
\nonumber
I(M)& = &\sum_{k=0}^M\mu(\phi_k;\KK)\log\mu(\phi_k;\KK)
(\phi_{k+1}-\phi_k)\\
& = &\frac{1}{M+1}\sum_{k=0}^M\mu(\phi_k;\KK)\log\mu(\phi_k;\KK) .
\end{eqnarray}
Identifying the probabilities 
\begin{equation}
\frac{\mu(\phi_k;\KK)}{M+1} = p_k(M)
\end{equation}
with the relative frequencies $p_k(M)$ of section \ref{ce} gives
\begin{equation}
I(M)=\sum_{k=0}^M p_k(M)\log[p_k(M)(M+1)]
\end{equation}
and thus 
\begin{equation}
I(M)=\log(M+1)(1-H^*(M))
\label{relation}
\end{equation}
provides a rigorous relationship between the entropies which becomes
exact in the limit $(L/a)\rai$. 

Note, however, that in the continuum limit $a\rao$ at fixed $L$
the normalization of $H^*(M)$ gives rise to a peculiar 
behaviour of this quantity.
While in this limit $\lim_{a\rao}I(M)=I(\KK)$ becomes a 
number depending on the local porosity fluctuations as
described by $\mu(\phi;\KK)$, equation (\ref{relation}) shows 
that $\lim_{a\rao}H^*(M)=1$ which is independent of the
microstructure. 
In other words the ``configuration'' entropy $H^*(M)$ should not
be used to distinguish different morphologies in the continuum limit.
We suggest calling $I(\KK)$ ``local geometry entropy''.
The generalized name accounts for the fact the $I(\KK)$ is 
readily generalized to morphological descriptors other than
porosity, such as local specific internal surface areas or local 
curvatures \cite{hil95d}.

\section{Entropy length}
\label{length}
An interesting quantity associated with local geometry distributions
is the entropy length.
The entropy length $L^*$ is defined as the length at which $I(M)$
becomes extremal, $\partial I(M)/\partial L|_{L=L^*}=0$ \cite{hil92f}.
Differentiating (\ref{relation}) yields
\begin{equation}
I^{'}(M)=\frac{1}{M+1}(1-H^*(M))-\log(M+1)H^{*'}(M)
\label{entlen}
\end{equation}
where the prime denotes the derivative with respect to $M$.
For large $M$, e.g. in the limit of high resolution $a\rao$
at fixed $L$, the first term  becomes negligible, and thus
the entropy length can be determined equally from the condition
$\partial H^*(M)/\partial L=0$.
This is illustrated in Figures \ref{pois} and \ref{comp} and \ref{lpent}.
Figure \ref{pois} shows the microstructure of a Poisson grain model 
obtained by placing circles of equal radii around centers 
which are distributed at random and with constant number density.
The corresponding entropy curves $I(M)$ and $H^*(M)$ are shown
in Figures \ref{comp} where, to facilitate the comparison, the curves 
have been shifted along the ordinate to have the same maximum 
value 0. We have assumed $a=1$ which allows one to plot the curves
as function of the size $L$ of the observation square.
Note that the curves show two extrema, of which the ones
at large $M$ coincide, while the extrema at small $M$ are 
different.
The entropy length 
is an accurate measure of the typical linear
size of the different phases, pores or components.

\section{Applications}

To illustrate the usefulness of the local geometry concept as
a morphological descriptor we apply it to several random 
microstructures.
We use two computer generated images of simple models for
disordered systems, and two experimentally obtained disordered
microstructures. The two synthetic images are generated from the
percolation model on a lattice and from the Poisson grain model 
consisting of uniformly distributed overlapping spheres.
The two experimental morphologies are observed when a thin
film of gold is deposited on a glass substrate, and when 
slicing through a natural sandstone.

\subsection{Simulated Bernoulli site percolation model}
Perhaps the simplest model of a random microstructure is the
site percolation model \cite{SA92}.
In this model, each lattice site has a probability 
$\langle \phi \rangle$ of being occupied and  
$(1 - \langle \phi \rangle)$ of being empty.
The occupation of sites is assumed to be
statistically independent.
A configuration of such a system in which the occupied
(black) sites represent pore space, and the unoccupied
ones matrix space is shown in Figure \ref{perc} for
$\langle \phi \rangle = 0.3$.

The local porosity distribution $\mu(\phi; {\KK})$ for the
percolation model depends only on $\langle \phi \rangle$ 
and $|\KK|=M=(L/a)^d$, and is given by the binomial distribution
\cite{hil96c,and96}.
If a hypercubic lattice with lattice constant $a$ is considered, 
then the LPD for a $d$-dimensional hypercubic measurement cell $\KK$ 
of side $L$ reads
\begin{eqnarray}
& & p_k(M) = \frac{\mu(\phi_{k};{\KK})}{M+1} =\\
& & \frac{M!}{\left[(M+1)\,\phi_{k}\right]!\:\left[(M+1)\,
\left(1-\phi_{k}-1/(M+1)\right)\right]!}\;
\langle\phi \rangle^{\,(M+1)\phi_{k}}\,(1-
\langle \phi \rangle )^{\,(M+1)(1-\phi_k-1/(M+1))}\nonumber
\end{eqnarray}
where $M=(L/a)^d$, $\phi_{k} = k/(M+1) \mbox{\ and\ } k= 0,1,\ldots,M$.
Its local geometry entropy
\begin{equation}
I(M) = \log(M+1)-H(M) = \log(M+1)+\sum_{k=0}^M p_k(M)\log(p_k(M))
\end{equation}
is displayed in Figures \ref{cent} and \ref{lpent} as the
curve with triangular symbols.
The extrema of the curve are at the boundaries.
This reflects correctly the fact that the microstructure
is homogenous and statistically independent at the microscopic
resolution (i.e. the lattice constant).
The value $I(1)=\log(2)+\langle\phi\rangle \log\langle\phi\rangle
+(1-\langle\phi\rangle)\log(1-\langle\phi\rangle)\approx 0.08228$
is approached for $M=1$. For $M\rightarrow \infty$ the entropy
$I(L)$ diverges to infinity (Note that Figure \ref{lpent} shows 
$-I(L)$ which diverges to $-\infty$), while $H^*(M)$ in
Figure \ref{cent} approaches a constant.

To demonstrate the influence of finite system size and statistics
we have plotted in Figures \ref{lpent} and \ref{cent} also the
curves $H^*(L)$ and $-I(L)$ determined directly from the image.
These curves are shown using circular symbols. The agreement with
the exact result is satisfactory although some deviations are apparent.

\subsection{Simulated Poisson grain model}
Another often used model for random microstructures is the
Poisson grain model.
In this model a constant number density of circles (spheres)
is randomly placed with uniform density into continuous space.
Figure \ref{pois} shows a random throw of disks with diameter
of 15 pixels in two dimensions. The porosity of the image
is $\langle \phi \rangle = 0.5$.

The entropy functions $-I(L)$ and $H^*(L)$ for this image
differ significantly from those for the percolation image.
Now the curves show a pronounced extremum while those for
the percolation case were monotonic.
Both curves exhibit an extremum at $L^*\approx 16$ corresponding
roughly to the circle diameter of 15 pixels. 
From other simulations we find that the extremum changes
with the circle (or sphere) diameter.
Hence the position of the extremum in $-I(L)$ or $H^*(L)$
is related to a characteristic length scale for the morphology.
This interpretation is also consistent for the percolation
model where the extremum occurs at $L^*\approx 1$ which
corresponds to the pixel diameter.

\subsection{Experimental gold film morphology}

The image of Figure \ref{gold} represents a digitized and
thresholded image of a transmission electron micrograph
of a thin film of gold on a glass substrate. 
The film was deposited by thermal evaporation under ultra-high
vacuum.
The evaporation was stopped before the gold phase,
shown in black, begins to percolate.
The surface coverage is $\langle \phi \rangle =0.413$.

The curves for the local porosity entropy and the configuration 
entropy of the gold film morphology from Figure \ref{gold} are
shown using square symbols in Figure \ref{lpent} and \ref{cent}.
Both curves exhibit again an extremum, this time at $L^*\approx 11$. 
This corresponds roughly to the characteristic thickness of
black and white regions in the image.
The length $L^*$ is expected to decrease down to the size of the elementary 
metallic grain at percolation \cite{and96}.
This is the main reason why in \cite{and96,ABL94} it was
suggested that the optical properties of these materials have to be
calculated at scale $L^*$.

\subsection{Experimental oolithic sandstone morphology}

In Figure \ref{sand} we show a slice through a natural
oolithic sandstone.
The image was obtained by slicing a sample of Savonnier
sandstone whose pore space was made visible by first
filling it with a coloured epoxy resin.
The cut surface was polished and photographed. 
The photgraph could then be digitized and thresholded
to give a black and white image.
The threshold was adjusted to match the measured
bulk porosity of $\langle \phi\rangle =0.186$.
In the image shown in Figure \ref{sand} the pore
space is coloured black while the rock matrix is
shown in white.

The entropy functions for the image of Figure \ref {sand}
are displayed as curves with cross symbols in
Figures \ref{lpent} and \ref{cent}.
Note that there is again a maximum in $-I(L)$ at
$L^*\approx 47$, but this maximum is much broader
and less pronounced than for the other images.
The curve for $H^*(L)$ in Figure \ref{cent}
also shows a maximum at $L^*\approx 55$.
It is so flat that it is difficult to distinguish
from the figure.

The difference between the values of $L^*$ obtained
from maximizing $H^*(L)$ as opposed to $-I(L)$ can be
understood from Eq. (\ref{entlen}).
The equality of the two entropy lengths is obtained
in the limit $M\rai$ in which the first term in
Eq. (\ref{entlen}) becomes negligible.
For finite $M$ the lengths can in general be different.
Hence we conclude that for the morphology of Figure
\ref{sand}, which represents the most heterogeneous case,
the resolution of the image needs to be improved before 
$I(M)$ and $H^*(M)$ will give the same entropy length.

\section{Discussion and Conclusion}
In this paper we have established an exact relationship between
the local porosity entropy \cite{hil92f} and the configurational
entropy of \cite{ABL94}.
These two morphological entropies were introduced independently 
by the authors to characterize random but correlated microstructures.
We have discussed the advantages and disadvantages of the two
entropies and suggest to call their common element local
geometry entropy.
It is argued that local geometry entropies are a useful
morphological indicator for random correlated morphologies.

The evaluation of the local geometry entropy for selected
morphologies indicates the existence of a minimum.
The position $L^*$ of the minimum in the entropy function is
called the entropy length. The length $L^*$ correlates
well with a characteristic length scale of the microstructure.
It is, however, different from the pixel-pixel correlation length.
Our analysis and past experience indicates that the entropy
length $L^*$, as measured here, corrresponds to that size of
measurement cells at which a finite fraction of measurement
cells begin to have vanishing local porosity.

Another observation concerns the width (or curvature) of
the extremum.
More heterogeneous microstructures such as the one in
Figure \ref{sand} appear to show a wider extremum than
those microstructures, such as Figures \ref{gold} or \ref{pois},
in which the black and white regions are more compact.
This observation is consistent with the findings in
\cite{hil92f} for synthetic morphologies.
Hence we conclude tentatively that the width of the
extremum correlates with the complexity or heterogeneity
of the microstructure.
More analyses of synthetic and experimental data are
desirable to further elucidate the geometrical and
morphological information content of the local
geometry concept.\\[3cm]
ACKNOWLEDGEMENT: We thank Joaquina Peiro, Patrice Gadenne, 
Christian Ostertag-Henning and Prof.Dr. R. Koch for providing
us with the experimental images. Some of us (R.H. and E.H.) 
are grateful to Norges Forskningsr{\aa}d for financial support.

\newpage
\begin{center}
FIGURE CAPTIONS
\end{center}
\begin{enumerate}
\item
\begin{enumerate}
\item {FIGURE \ref{perc}}\\
Random microstructure of the (uncorrelated) site
percolation model in which occupied (black) points
represent pore space. The porosity (volume fraction
of pore space) in the image is $\langle \phi \rangle = 0.3$.
\label{perc}
\item {FIGURE \ref{pois}}\\
Poisson grain model configuration with constant point density
of circles. Total volume fraction of $\langle \phi \rangle = 0.5$
\label{pois}
\item {FIGURE \ref{gold}}\\
Discontinuous thin film morphology of gold at volume fraction
$\langle \phi \rangle = .413$
\label{gold}
\item {FIGURE \ref{sand}}\\
Planar thin section through a Savonnier oolithic
sandstone with pore space rendered black.
The side length of the image corresponds to 
roughly 1 cm, the total porosity is
$\langle \phi \rangle = .186$
\label{sand}
\end{enumerate}
\item {FIGURE \ref{comp}}\\
Comparison of local porosity entropy $-I(L)$ and
configuration entropy $H^*(L)$ for the Poisson 
grain model shown in Figure \protect\ref{pois} with 
disks diameter 15 pixels. 
Note that the curves show extrema at the same
length $L\approx 26$.
The curves are shifted to have the same ordinate
$-I(L)=H^*(L)=0$ at the extremum.
\label{comp}
\item {FIGURE \ref{lpent}}\\
Local porosity entropy $-I(L)$ \protect\cite{hil92f} as a function of the
side length of the measurement cell calculated for
all random microstructures shown in Figures \protect\ref{perc},
\protect\ref{pois}, \protect\ref{gold} and \protect\ref{sand}. 
The circles correspond to the site percolation image 
shown in Figure \protect\ref{perc}. The triangles are the
theoretical values for an infinitely large image at
the same bulk porosity.
The diamonds are the result for the Poisson grain model 
shown in Figure \protect\ref{pois}.
The squares represent the experimentally observed
gold film morphology of Figure \protect\ref{gold}, while 
the crosses correspond to the sandstone cross
section shown in Figure \protect\ref{sand}.
\label{lpent}
\item {FIGURE \ref{cent}}\\
Configuration entropy $H^*(L)$ \protect\cite{BALPP93,ABL94} as a function 
of the side length of the measurement cell calculated for
all random microstructures shown in Figures \protect\ref{perc},
\protect\ref{pois}, \protect\ref{gold} and \protect\ref{sand}. 
The symbol usage is the same as in Figure \protect\ref{lpent}
and is explained in the figure caption there.
\label{cent}
\end{enumerate}

\newpage

\bibliographystyle{siam}
\bibliography{$HOME/tex/bib/porous,$HOME/tex/bib/publ,lafait}
\end{document}